\documentclass[aps,pra,twocolumn,superscriptaddress,showpacs]{revtex4}
\usepackage{graphicx}
\usepackage{epsfig}
\usepackage{amsmath}
\usepackage{amssymb}
\usepackage{amsfonts}
\usepackage{mathrsfs}
\usepackage{theorem}
\usepackage{bm}
\usepackage{url}
\usepackage[T1]{fontenc}
\usepackage{csquotes}
\usepackage{natbib}
\usepackage{float}
\usepackage{appendix}
\MakeOuterQuote{"}

\usepackage{dcolumn}
\usepackage{color}

\begin{document}

\title{Experimental realisation of generalised qubit measurements based on quantum walks}

\author{Yuan-yuan Zhao}
\affiliation{Key Laboratory of Quantum Information, University of Science and Technology of China, CAS, Hefei, 230026, People's Republic of China}
\affiliation{Synergetic Innovation Center of Quantum Information and Quantum Physics, University of Science and Technology of China, Hefei, Anhui 230026, People's Republic of China}
\author{Neng-kun Yu}
\affiliation{The Institute for Quantum Computing, University of Waterloo, Waterloo, Ontario, Canada}
\affiliation{Department of Mathematics\&Statistics, University of Guelph, Guelph, Ontario, Canada}
\author{Pawe{\l} Kurzy\'{n}ski}
\affiliation{Centre for Quantum Technologies, National University of Singapore, 3 Science Drive 2, 117543 Singapore, Singapore}
\affiliation{Faculty of Physics, Adam Mickiewicz University, Umultowska 85, 61-614 Pozna\'{n} , Poland}
\author{Guo-yong Xiang}
\email{gyxiang@ustc.edu.cn}
\affiliation{Key Laboratory of Quantum Information, University of Science and Technology of China, CAS, Hefei, 230026, People's Republic of China}
\affiliation{Synergetic Innovation Center of Quantum Information and Quantum Physics, University of Science and Technology of China, Hefei, Anhui 230026, People's Republic of China}
\author{Chuan-Feng Li}
\affiliation{Key Laboratory of Quantum Information, University of Science and Technology of China, CAS, Hefei, 230026, People's Republic of China}
\affiliation{Synergetic Innovation Center of Quantum Information and Quantum Physics, University of Science and Technology of China, Hefei, Anhui 230026, People's Republic of China}

\author{Guang-Can Guo}
\affiliation{Key Laboratory of Quantum Information, University of Science and Technology of China, CAS, Hefei, 230026, People's Republic of China}
\affiliation{Synergetic Innovation Center of Quantum Information and Quantum Physics, University of Science and Technology of China, Hefei, Anhui 230026, People's Republic of China}

\begin{abstract}
We report an experimental implementation of a single-qubit generalised measurement scenario (POVM) based on a quantum walk model. The qubit is encoded in a single-photon polarisation. The photon performs a quantum walk on an array of optical elements, where the polarisation-dependent translation is performed via birefringent beam displacers and a change of the polarisation is implemented with the help of wave-plates. We implement: (i) Trine-POVM, i.e., the POVM elements uniformly distributed on an equatorial plane of the Bloch sphere; (ii) Symmetric-Informationally-Complete (SIC) POVM; and (iii) Unambiguous Discrimination of two non-orthogonal qubit states.
\end{abstract}
\date{\today }

\maketitle

\section{Introduction}
The basic unit of quantum information is a two-level quantum system commonly known as a qubit. Qubits can be implemented on physical objects such as polarisation of photons or intrinsic angular momentum (spin $1/2$) of quantum particles. Any quantum computation relies on precise preparations, transformations and measurements of such systems. Before actual quantum computer is build, one has to master the ability to manipulate with single qubits and learn how to readout information encoded in them.

The information readout from a quantum system is done via a measurement. In most common scenario one performs the von Neumann measurement that projects a state of the qubit onto one of two perfectly distinguishable (orthogonal) physical states of the system. Such measurements are sharp in a sense that once the measurement is done, the outcome of the measurement is determined and any repetition of exactly the same measurement would yield the same outcome value.

Physically, von Neumann measurements are realised via interaction of the system with the measurement apparatus. The pointer of the measurement apparatus is represented via wave packed and the interaction causes this wave packet to move either to the left or right, depending on the value of the measured observable. In general, this value might be undetermined and the pointer goes into superposition of being to the left and to the right from its initial position. The actual collapse of wave function is usually attributed to the observer who reads out the measurement outcome from the pointer. The sharpness of the measurement comes from the fact that the initial spread of the pointer's state is assumed to be narrow and the translation caused by the interaction with the measured observable is large enough to prevent overlap between the part of the wave packed that was shifted to the right with the part that was shifted to the left.

On the other hand, there are generalised measurement scenarios, the so called Positive-Operator Valued Measure (POVM), in which one allows the principal system to interact with an ancillary system, whose state is known, and later performs a von Neumann measurement on the joint system. This effectively extends the dimensionality of the Hilbert space and one can implement measurements of a quibit with more than two outcomes. As a result, one gains a plethora of new possibilities. They allow one, for example, to perform a tomography of qubit with a single measurement setup \cite{SIC1,SIC2}, or to discriminate between non-orthogonal quantum states \cite{SD1,SD2}. POVMs were implemented in laboratories using various setups \cite{POVM1,POVM2,POVM3,POVM4,POVM5,TSExp,DisExp}. 

In \cite{QWPOVM} it was proposed that they can be implemented via a discrete-time quantum walk which has been  realised in a laboratory using various physical systems \cite{Exp1,Exp2,Exp3,Exp4,Exp5,Exp6,Exp7,Exp8,Exp9,Exp10,Exp11,Exp12,AWhite}. Quantum walks model an evolution of a particle in a discrete space. The particle moves either one step to the left or right, depending on a state of a two-level system known as a coin. Quantum walks were originally proposed as an interference process resulting from a modified version of a von Neumann measurement in which a pointer state distribution is much broader than the shift of its position \cite{Aharonov}. The position of the pointer plays the role of the quantum walker and the qubit that is measured plays the role of the coin. If the interaction between the pointer and the qubit occurs at the same time as the evolution of the qubit, the measured value changes during the process and the pointer starts to move back and forth. This movement leads to an interference and the interference pattern produced in this process can be interpreted as a POVM. In fact, in \cite{QWPOVM} it was shown that any POVM can be implemented in such a way, provided a necessary evolution is applied to the qubit.

In this work we report an experimental implementation of the above quantum walk POVM scenario. Here, we use the optical setup in which the qubit is encoded in a polarization state of a single photon and the position of the quantum walker is implemented on a photonic path \cite{AWhite}. We construct setups realising (i) three POVM elements symmetrically distributed on an equatorial plane of the Bloch sphere (Trine-POVM); (ii) Symmetric-Informationally-Complete (SIC) POVM; and (iii) Unambiguous Discrimination of two non-orthogonal quantum states.

\section{Discrete-time quantum walks}

Discrete-time quantum walks are quantum counterparts of classical random walks in which a particle takes a random step to the left or right. In the classical case the particle spreads in a diffusive manner (a standard deviation of its position is proportional to $\sqrt{t}$) and after many steps a spatial probability distribution is Gaussian. In quantum case the system is described by two degrees of freedom $|\psi\rangle=|x,c\rangle$: the position $x=\dots,-1,0,1,\dots$ and a two-level internal degree of freedom known as a coin $c=\leftarrow,\rightarrow$. The evolution is unitary and consists of two sub-operations $U=CT$, namely a unitary coin toss (that is a $2\times 2$ unitary matrix acting only on the coin degree of freedom)
\begin{equation}\label{C}
C=\begin{pmatrix} \cos\theta & e^{-i\beta}\sin\theta \\ -e^{i\beta}\sin\theta & \cos\theta \end{pmatrix}
\end{equation}
and a conditional translation operation
\begin{equation}\label{T}
T=\sum_x |x+1,\rightarrow\rangle \langle x,\rightarrow | + |x-1,\leftarrow \rangle \langle x,\leftarrow |.
\end{equation}

Quantum walks with uniform (position-independent) coin operation spread ballistically (standard deviation proportional to $t$) and its probability distribution differs form the classical Gaussian shape \cite{QW1,QW2}. On the other hand, quantum walks with position-dependent coin operation
\begin{equation}\label{C_x}
C_{x}=\begin{pmatrix} \cos\theta_x & e^{-i\beta_x}\sin\theta_x \\ -e^{i\beta_x}\sin\theta_x & \cos\theta_x \end{pmatrix}
\end{equation}
can be used to observe localisation \cite{Loc1,Loc2}, or to simulate physical systems with non-homogenous interactions \cite{Transf}. In \cite{QWPOVM} it was shown that quantum walks with position and time-dependent coin operations $C_{x,t}$ can be also used to implement POVMs.

\section{POVM implementation with quantum walks}

The probability distribution of a quantum walk, that is initially localised at the origin $x=0$, depends on the initial coin state $\alpha|\rightarrow\rangle + \beta|\leftarrow\rangle$ and on the subsequent coin operations. A single application of the operator $T$ makes the particle to go into superposition $\alpha|1,\rightarrow\rangle + \beta|-1,\leftarrow\rangle$. In this case the position measurement would correspond to a von Neumann measurement of the coin in the basis $\{\leftarrow,\rightarrow\}$, where the result $\rightarrow$ corresponds to finding the particle at $x=1$ and $\leftarrow$ to finding it at $x=-1$. However, if one allows quantum walk to evolve for more than one step and the coin operation to change from one step to another step, then the particle can spread over many positions and the measurement of its location at $x$ may correspond to a measurement of some POVM element $E_x$. Indeed, a proper choice of coin operations can lead to an arbitrary qubit POVM scenario \cite{QWPOVM}.

The POVM elements $E_i$ ($i=1,2,\dots,n$) are nonnegative operators obeying
\begin{equation}
\sum_i^n E_i = \openone.
\end{equation}
They differ from standard von Neumann projectors $\Pi_i$ in that they do not have to be orthonormal ($\Pi_i \Pi_j = \delta_{i,j}\Pi_i$, but $E_i E_j \neq \delta_{i,j}$) and that their number can be greater than the dimension of the system $n>d$. The following quantum walk algorithm, proposed in \cite{QWPOVM}, generates an arbitrary set of rank 1 POVM elements $\{E_1,\dots,E_n\}$ (rank 2 elements can be generated with a modified version of this algorithm).
\begin{enumerate}
\item Initiate the quantum walk at position $x=0$ with the coin state corresponding to the qubit state one wants to measure.
\item Set $i := 1$.
\item While $i < n$ do the following:
	\begin{itemize}
	\item Apply coin operation $C_i^{(1)}$ at position $x = 0$ and identity elsewhere and
then apply translation operator $T$.
	\item Apply coin operation $C_i^{(2)}$ at position $x = 1$, $$NOT = \begin{pmatrix} 0 & 1 \\
1 & 0 \end{pmatrix}$$ at position $x = -1$ and identity elsewhere and then apply
translation operator $T$.
	\item  $i := i + 1$.
	\end{itemize}
\end{enumerate}
Since in this work we are only interested in the measurement statistics, and not in the post-measurement states, we simplified the algorithm in \cite{QWPOVM} and omitted the last step. The POVM elements that are generated depend solely on the form of operators $C_i^{(1)}$ and $C_i^{(2)}$.

\section{Experiment}

In our experiment, frequency-doubled femtosecond pulse (390 nm, 76 MHz repetition rate, 80 mw average power) from a mode-locked Ti:sapphire laser pump a type-I beta-barium-borate (BBO, 6.0$\times$6.0$\times$2.0 mm$^{3}$, $\theta$= 29.9) crystal to produces the degenerate photon pairs. After being redirected by the mirrors (M1 and M2, as in the Fig. 1(a)) and the interference filters (IF, $\bigtriangleup\lambda$=3 nm, $\lambda$=780 nm), the photon pairs generated in the spontaneous parametric down-conversion (SPDC) process are coupled into single-mode fibers separately. Single photon state is prepared by triggering on one of these two photons, and the coincidence counting rate collected by the avalanche photo-diodes (APD) are about $4\times10^{4}$ in one minute.

\begin{figure}[tbph]
\begin{center}
\includegraphics [width= 0.9 \columnwidth]{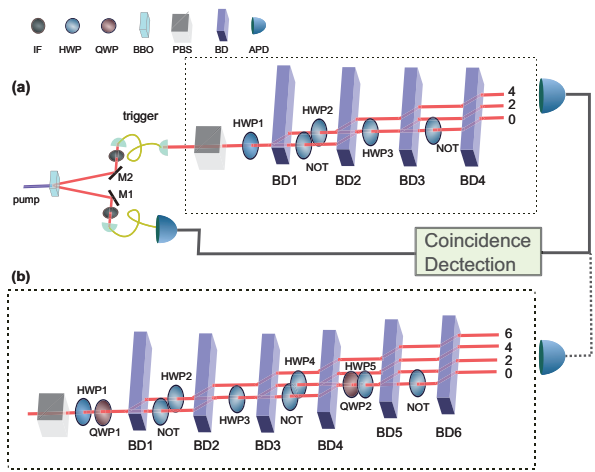}
\end{center}
\caption{Experimental setup. (a) Experimental setup for constructing the trine POVMs corresponding to $|\psi_3^{i}\rangle$ and realizing the unambiguous state discrimination. (b) Optical network for constructing SIC-POVMs. Initial coin states are prepared by passing the single photons through a polarizing beam splitter (PBS), a half-wave plate (HWP) HWP1 and a quarter-wave plate (QWP) QWP1 in a specific configuration. The conditional position shifts are implemented by Beam Displacers (BDs) and the coin operators in different positions are realized by wave plates with different angles (Table 1). The indices in the figure denote the position of walker.
}
\label{fig:map}
\end{figure}

One-dimensional discrete time quantum walk system has two degrees of freedom, $x$ and $c$, $x$ is the position of the particle and c is the state of the coin. In our experiment they are encoded in the longitudinal
spatial modes and polarizations {$|H\rangle$, $|V\rangle$} of the single photons respectively. In this case, the conditional translation operator as given by Eq.(2) is realized by the designed BD, that does not displace the vertical polarized photons ($|x, V\rangle\rightarrow|x-1, V\rangle$) but makes the horizontal polarized ones undergo a 4 mm lateral displacement ($|x, H
  \rangle\rightarrow|x+1, H\rangle$).

\subsection{Trine POVM}
 The experimental setup in Fig. 1(a) is used to construct the trine POVMs, $\frac{2}{3}|\psi_{3}^{i}\rangle\langle\psi_{3}^{i}| (i=1, 2, 3)$,
\begin{equation}
\begin{aligned}
   &|\psi_{3}^{1}\rangle=|H\rangle
   \\&|\psi_{3}^{2}\rangle=-\frac{1}{2}(|H\rangle-\sqrt{3}|V\rangle)
   \\&|\psi_{3}^{3}\rangle=-\frac{1}{2}(|H\rangle+\sqrt{3}|V\rangle).
\end{aligned}
\end{equation}
  According to the settings of the coin operators, the optical axes of BD1 and BD2 must be aligned, in other words, they form an interferometer. When rotating HWP1 and HWP3 by $22.5^{\circ}$, we observed that the interference visibility of the interferometer was about $99.8$ and the system was stable over 2.5 hours of timescale. After aligning, we begin to set the corresponding coin operators in each step. For the Trine POVMs, we have
  \begin{equation}\label{C_trine}
  \begin{aligned}
    &C_{1}^{(1)}=1, C_{1}^{(2)}=\sqrt{\frac{1}{3}}\begin{pmatrix}\sqrt{2} & 1 \\ 1 & -\sqrt{2} \end{pmatrix},
  \\&C_{2}^{(1)}=\sqrt{\frac{1}{2}}\begin{pmatrix}1 & 1 \\ 1 & -1 \end{pmatrix}, C_{2}^{(2)}=1,
  \end{aligned}
  \end{equation}
  where $C_{1}^{(2)}$ and $C_{2}^{(1)}$ are realized by rotating HWP2 and HWP3 by $17.32^{\circ}$ and $22.5^{\circ}$ respectively. The initial trine coin states $|\psi_{3}^{i}\rangle$ are constructed by rotating HWP1 by $0^{\circ}$, $-30^{\circ}$ and $30^{\circ}$, while the anti-trine states $|\bar{\psi}_{3}^{i}\rangle$ are constructed by rotating it by $45^{\circ}$, $15^{\circ}$ and $-15^{\circ}$. At last, every output port's detection efficiency are calibrated so that the differences among them are below $5\%$.

\begin{figure}[tbph]
\begin{center}
\includegraphics [width=7.5cm,height=9cm]{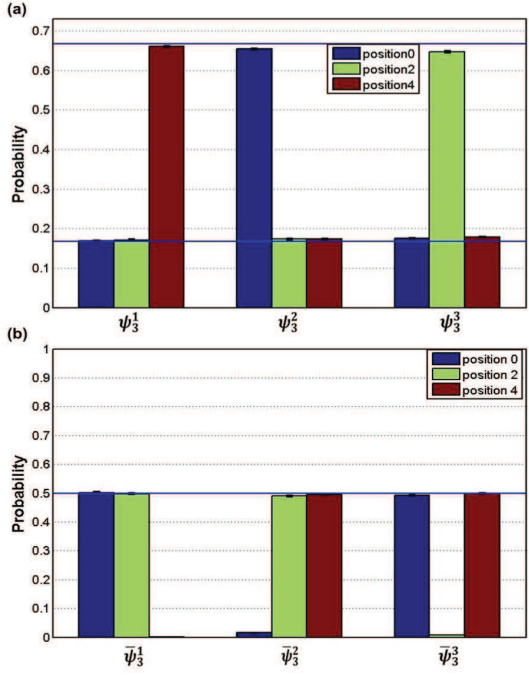}
\end{center}
\caption{Results for trine POVMs. Histogram shows the probabilities of counting rates in position 0, 2 and 4 with input states $\psi_{3}^{i}$(a) and $\bar{\psi}_{3}^{i}$(b), respectively. All results are normalized so that the sum of the counts in these three positions is 1. The theoretical values are shown as the blue lines, which are 2/3, 1/6, 1/6 for $\psi_{3}^{i}$ and 1/2, 1/2, 0 for $\bar{\psi}_{3}^{i}$(i=1, 2, 3); error bars are too small to identify.}
\label{fig:3POVM}
\end{figure}

 Fig. 2 shows that the results in our experiment agree with theoretical predictions. The ratios $2/3:1/6:1/6$ ($0:1/2:1/2$) for the cases of $|\psi_{3}^{i}\rangle$ ($|\bar{\psi}_{3}^{i}\rangle$, where $\langle\psi_{3}^{i}|\bar{\psi}_{3}^{i}\rangle=0$) are given in theory and the detailed numerical results of the probability distributions can be found in Table II and III. To visualize that the setup has constructed the Trine POVMs, it is important to demonstrate that we cannot find states $|\bar{\psi}_{3}^{1}\rangle$ in position 4, $|\bar{\psi}_{3}^{2}\rangle$ in position 0 and $|\bar{\psi}_{3}^{3}\rangle$ in position 2. Fig. 2(b) and Table III show that the probabilities of these events are indeed very close to zero, with an average value of $0.0085$.
 In addition, the results for states $|\psi_{3}^{i}\rangle$ also indicate the coefficients of the POVM we constructed is $\frac{2}{3}$, see Fig. 2(a). The errors in our experiment mainly stem from the imperfect wave plates and the interferometers and the counting statistics of the photons.

 \subsection{SIC POVM}
 The optical network in Fig. 1(b) construct the SIC POVMs, $\frac{1}{2}|\psi_{4}^{i}\rangle\langle\psi_{4}^{i}| (i=1, 2, 3, 4)$,
\begin{equation}
\begin{aligned}
   &|\psi_{4}^{1}\rangle=|H\rangle
   \\&|\psi_{4}^{2}\rangle=-\frac{1}{\sqrt{3}}|H\rangle+\sqrt{\frac{2}{3}}|V\rangle
   \\&|\psi_{4}^{3}\rangle=-\frac{1}{\sqrt{3}}|H\rangle+e^{{i\frac{2\pi}{3}}}\sqrt{\frac{2}{3}}|V\rangle
   \\&|\psi_{4}^{4}\rangle=-\frac{1}{\sqrt{3}}|H\rangle+e^{{-i\frac{2\pi}{3}}}\sqrt{\frac{2}{3}}|V\rangle.
\end{aligned}
\end{equation}
 The coin operators
 \begin{equation}\label{C_SIC}
  \begin{aligned}
    &C_{1}^{(1)}=1, C_{1}^{(2)}=\frac{1}{\sqrt{2}}\begin{pmatrix}-1 & 1 \\ 1 & 1 \end{pmatrix},
  \\&C_{2}^{(1)}=\frac{1}{\sqrt{2}}\begin{pmatrix}-1 & 1 \\ 1 & 1 \end{pmatrix}, C_{2}^{(2)}=\frac{1}{\sqrt{3}}\begin{pmatrix}\sqrt{2} & 1 \\ 1 & -\sqrt{2} \end{pmatrix},
  \\&C_{3}^{(1)}=\frac{1}{\sqrt{2}}\begin{pmatrix}e^{-i\frac{\pi}{3}} & e^{i\frac{\pi}{6}}\\ e^{i\frac{\pi}{3}} & e^{-i\frac{\pi}{6}} \end{pmatrix}, C_{3}^{(2)}=1
  \end{aligned}
  \end{equation}
  are realized by wave plates in various configurations (details in Table I).
 \begin{table}[H]
$$
\begin{array}{|c|c|c|c|c|c|}
    \hline
          &  C_{1}^{2}       & C_{2}^{1}        & C_{2}^{2}         & C_{3}^{1}           & C_{3}^{2}\\
    \hline
    HWP   & 67^{\circ}30^{'} & 67^{\circ}30^{'} & 17^{\circ}38^{'}  &142^{\circ}30^{'}    & --  \\
    \hline
    QWP   &   --             &     --           &     --            & 150^{\circ}         &--\\
    \hline
\end{array}
$$
\caption{The configurations of the QWPs and HWPs to realize the coin operators for constructing the SIC-POVM.}
\end{table}
For these settings, BD1 and BD2, BD3 and BD4 form two interferometers whose interference visibility are both above 0.993. The HWP1 and QWP1 with different angles in front of a polarizing beam splitter (PBS) are used to produce
$\psi_{4}^{i}$ and the corresponding orthogonal states $\bar{\psi}_{4}^{i}$ (Table VI and Table VII). As shown in Fig. 3, the results are also in good accordance with theoretical ratios $\frac{1}{2}:\frac{1}{6}:\frac{1}{6}:\frac{1}{6}$ for $\psi_{4}^{i}$ and $\frac{1}{3}:\frac{1}{3}:\frac{1}{3}:0$ for $\bar{\psi}_{4}^{i}$.

\begin{figure}[tbph]
\begin{center}
\includegraphics [width=7.5cm,height=9cm]{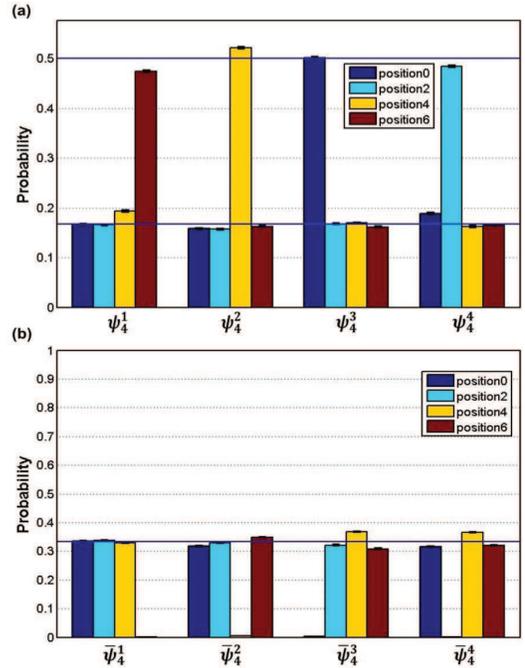}
\end{center}
\caption{Results for SIC POVMs. Histogram showing the normalized probability of counting rate in position 0, 2, 4 and 6 respectively with the input state $\psi_{4}$(a) and $\bar{\psi}_{4}^{i}$(b). The theoretical values are shown as the blue lines, which are given by 1/2, 1/6, 1/6, 1/6 for states $\psi_{4}^{i}$ and 1/3, 1/3, 1/3, 0 for states $\bar{\psi}_{4}$; error bars are too small to identify.}
\label{fig:4POVM}
\end{figure}

 \subsection{Unambiguous state discrimination}
\begin{figure}[tbph]
\begin{center}
\includegraphics [width= 0.9 \columnwidth]{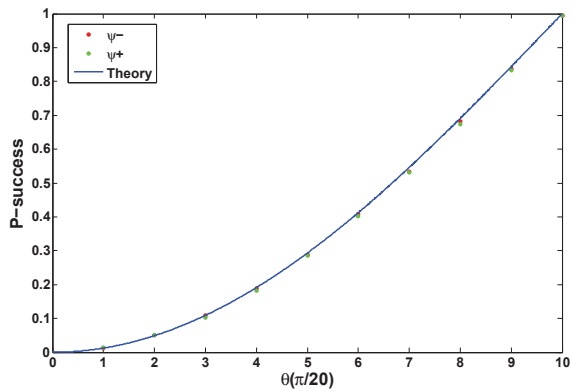}
\end{center}
\caption{Theoretical and experimental successful probability v.s. $\theta$ which is related to the state to be discrimination; error bars are too small to identify.}
\label{fig:2POVM}
\end{figure}

For the unambiguous state discrimination of states,
$|\psi_{\pm}\rangle=\cos(\theta/2)|H\rangle\pm\sin(\theta/2)|V\rangle$,
we can use the same setup as in Fig. 1(a), with
\begin{equation}\label{C_2state}
  \begin{aligned}
    &C_{1}^{(1)}=1,
    \\&C_{1}^{(2)}=\begin{pmatrix}\sqrt{1-(\tan{\frac{\theta}{2}})^2} & \tan{\frac{\theta}{2}} \\ \tan{\frac{\theta}{2}} & -\sqrt{1-(\tan{\frac{\theta}{2}})^2}\end{pmatrix},
  \\&C_{2}^{(1)}=\sqrt{\frac{1}{2}}\begin{pmatrix}1 & 1 \\ 1 & -1 \end{pmatrix}, C_{2}^{(2)}=1,
  \end{aligned}
\end{equation}

The state after four quantum walk steps becomes 
\begin{equation}
|\psi_{+}\rangle=\sqrt{\cos\theta}|4,H\rangle+\sqrt{2}\sin{\frac{\theta}{2}}|2,H\rangle, 
\end{equation}
or
\begin{equation}
|\psi_{-}\rangle=\sqrt{\cos\theta}|4,H\rangle-\sqrt{2}\sin{\frac{\theta}{2}}|0,H\rangle. 
\end{equation}
Therefor, if the photon is detected at position $x=0$ one knows that the coin was definitely in the state $|\psi_{-}\rangle$. If it was detected at position $x=2$ one knows that the coin was definitely in the state $|\psi_{+}\rangle$. Finally,  if it was detected at position $x=4$ one gains no information and the discrimination was unsuccessful.

In our experiment, the input states $\psi_{+}$ and $\psi_{-}$ are prepared in $\frac{\pi}{20}$ steps from $0\leq\theta\leq\frac{\pi}{2}$. The states for various $\theta$ are prepared by rotating the HWP, placed before the polarizing beam splitter (PBS), and the coin operator $C_{1}^{(2)}$, $C_{2}^{(1)}$and the NOT operator are realized by a HWP rotating in the angle of $\frac{1}{2}\arcsin(\tan{\frac{\theta}{2}})$, $\frac{\pi}{8}$ and $\frac{\pi}{4}$ respectively. From Fig. 4 we can see that the probability of the successful discrimination is increasing with $\theta$. For more details see Tables VIII and IX.

\section{Conclusions}

We experimentally realised three generalised measurement scenarios for a qubit. These scenarios are based on a quantum walk model presented in \cite{QWPOVM}. Our results match the theoretical predictions. We believe that these kind of experimental setups can be used in the future to implement other types of generalised measurement scenarios with multiple outcomes and rank 2 POVM elements, and to study quantum walks with position and time-dependent coin operations. Finally, we would like to mention that similar results had been reported while we were working on this experiment \cite{Xue1,Xue2}.

\section{Acknowledgements}

The authors would like to thank Yongsheng Zhang for helpful discussion. The work in USTC is supported by National Fundamental Research Program (Grants No. 2011CBA00200 and No. 2011CB9211200), National Natural Science Foundation of China (Grants No. 61108009 and No. 61222504). P.K. is supported by the National Research Foundation and Ministry of Education in Singapore.

\section{Appendix}
\appendix
The detailed results about $\psi_{3}^{i}$, $\bar{\psi}_{3}^{i}$, $\psi_{4}^{i}$ and $\bar{\psi}_{4}^{i}$ are shown in Table II, Table III, Table IV and Table V. Table VI and Table VII are the angles of the HWP1 and QWP1 for preparing states $\psi_{4}^{i}$ and $\bar{\psi}_{4}^{i}$. The main parameters of unambiguous state discrimination and the detailed results can be found in Table VIII and Table IX.


\begin{table}[H]
$$
\begin{array}{|c|c|c|c|c|c|}
    \hline
    state          &  P_{0}           &P_{2}             & P_{4}            \\
    \hline
    \psi_{3}^{1}   & 0.1684(20)  & 0.1711(19)  & 0.6604(25)  \\
    \hline
    \psi_{3}^{2}   & 0.6540(26)  & 0.1731(20)  & 0.1729(21)  \\
    \hline
    \psi_{3}^{3}   & 0.1753(21)  & 0.6466(25)  & 0.1782(21) \\
    \hline

\end{array}
$$
\caption{\footnotesize{$P_{0}$, $P_{2}$ and $P_{4}$ are the nomarized probabilities of counting rate in position 0, 2 and 4. }}
\end{table}

\renewcommand\arraystretch{1.5}
\begin{table}[H]
$$
\begin{array}{|c|c|c|c|c|c|}
    \hline
    state                 &  P_{0}           &P_{2}             & P_{4}            \\
    \hline
    \bar{\psi}_{3}^{1}    & 0.5018(27)  & 0.4977(27)  & 0.0005(01)  \\
    \hline
    \bar{\psi}_{3}^{2}    & 0.0151(06)  & 0.4897(27)  & 0.4952(26)  \\
    \hline
    \bar{\psi}_{3}^{3}    & 0.4933(25)  & 0.0081(05)  & 0.4987(25) \\
    \hline

\end{array}
$$
\caption{\footnotesize{$P_{0}$, $P_{2}$ and $P_{4}$ are the nomarized probabilities of counting rate in position 0, 2 and 4. }}
\end{table}

\begin{table}[H]
$$
\begin{array}{|c|c|c|c|c|c|}
    \hline
      state          &  P_{0}           &P_{2}             & P_{4}     & P_{6}                    \\
    \hline
    \psi_{4}^{1}     & 0.1662(19)  & 0.1654(19)  & 0.1934(20) & 0.4749(26)    \\
    \hline
    \psi_{4}^{2}     & 0.1585(18)  & 0.1571(18)  & 0.5220(27) & 0.1625(20)    \\
    \hline
    \psi_{4}^{3}     & 0.5015(26)  & 0.1676(19)  & 0.1695(19) & 0.1614(19)    \\
    \hline
    \psi_{4}^{4}     & 0.1885(20)  & 0.4843(25)  & 0.1623(19) & 0.1649(19)    \\
    \hline

\end{array}
$$
\caption{\footnotesize{The nomarized probabilities of counting rates for state $\psi_{4}^{i}$ in position 0, 2, 4 and 6. }}
\end{table}

\renewcommand\arraystretch{1.5}
\begin{table}[H]
$$
\begin{array}{|c|c|c|c|c|c|}
    \hline
      state          &  P_{0}           &P_{2}             & P_{4}     & P_{6}                     \\
    \hline
    \bar{\psi}_{4}^{1}     & 0.3341(24)  & 0.3367(25)  & 0.3289(24) & 0.0003(01)    \\
    \hline
    \bar{\psi}_{4}^{2}     & 0.3182(23)  & 0.3283(24)  & 0.0051(04) & 0.3485(24)    \\
    \hline
    \bar{\psi}_{4}^{3}     & 0.0040(03)  & 0.3209(24)  & 0.3671(25) & 0.3080(23)    \\
    \hline
    \bar{\psi}_{4}^{4}     & 0.3152(24) & 0.0005(01)  & 0.3647(24) & 0.3196(24)    \\
    \hline

\end{array}
$$
\caption{\footnotesize{The nomarized probabilities of counting rates for state $\bar{\psi}_{4}^{i}$ in position 0, 2, 4 and 6. }}
\end{table}

\renewcommand\arraystretch{1.5}
\begin{table}[H]
$$
\begin{array}{|c|c|c|c|c|c|}
    \hline
          &  \psi_{4}^{1} & \psi_{4}^{2} & \psi_{4}^{3} & \psi_{4}^{4}  \\
    \hline
    HWP1   & 0^{\circ} & -27^{\circ}22^{'} & 17^{\circ}38^{'}  &45^{\circ}   \\
    \hline
    QWP1   & 0^{\circ} & 35^{\circ}16^{'}  & -27^{\circ}22^{'} & -27^{\circ}22^{'}\\
    \hline
\end{array}
$$
\caption{The angles of HWP1 and QWP1 used to prepare the states $\psi_{4}^{i}$.}
\end{table}

\renewcommand\arraystretch{1.5}
\begin{table}[H]
$$
\begin{array}{|c|c|c|c|c|c|}
    \hline
          &\bar{\psi}_{4}^{1} & \bar{\psi}_{4}^{2} & \bar{\psi}_{4}^{3} & \bar{\psi}_{4}^{4} \\
    \hline
    HWP1   &45^{\circ} & 17^{\circ}38^{'} & -27^{\circ}22^{'} & 0^{\circ}\\
    \hline
    QWP1   &0^{\circ} & 35^{\circ}16^{'}  & -27^{\circ}22^{'} & -27^{\circ}22^{'}\\
    \hline
\end{array}
$$
\caption{The angles of HWP1 and QWP1 used to prepare the states $\bar{\psi}_{4}^{i}$.}
\end{table}

\begin{table}[H]
$$
\begin{array}{|c|c|c|c|c|c|}
    \hline
    \theta  & \theta_{1/2} & C_{1}^{(2)}  &P_{theory} & P_{e}                     \\
    \hline
    \pi/20  & 2^{\circ}15^{'}  & 2^{\circ}15^{'}  & 0.0123  & 0.013(06)  \\
    \hline
    \pi/10  & 4^{\circ}30^{'}  & 4^{\circ}34^{'}  & 0.0489  & 0.050(10)   \\
    \hline
    3\pi/20 & 6^{\circ}45^{'}  & 6^{\circ}57^{'}  & 0.0109  & 0.103(15)  \\
    \hline
    \pi/5   & 9^{\circ}        & 9^{\circ}29^{'}  & 0.191   & 0.181(19)  \\
    \hline
    \pi/4   & 11^{\circ}15^{'} & 12^{\circ}14^{'} & 0.293   & 0.285(22)  \\
    \hline
    3\pi/10 & 13^{\circ}30^{'} & 15^{\circ}19^{'} & 0.412   & 0.402(23)  \\
    \hline
    7\pi/20 & 15^{\circ}45^{'} & 18^{\circ}54^{'} & 0.546   & 0.531(24)  \\
    \hline
    2\pi/5  & 18^{\circ}       & 23^{\circ}18^{'} & 0.691   & 0.673(22)  \\
    \hline
    9\pi/20 & 20^{\circ}15^{'} & 29^{\circ}21^{'} & 0.844   & 0.832(18) \\
    \hline
    \pi/2   & 22^{\circ}30^{'} & 45^{\circ}       & 1.000   & 0.996(03)  \\
    \hline
\end{array}
$$
\caption{\footnotesize{$\theta $, the angle related to the input state $\psi_{+}$ which is prepared by HWP1 rotated to $\theta_{1/2}$; $C_{1}^2$, the angle of HWP2 to realize the operator $C_{1}^2$. $P_{theory}$ and $P_{e}$ represent the theoretical and the experimental successful probability. }}
\end{table}

\begin{table}[H]
$$
\begin{array}{|c|c|c|c|c|c|}
    \hline
    \theta   & \theta_{1/2}      & C_{1}^{(2)}          &P_{theory} & P_{e} \\
    \hline
    -\pi/20  & -2^{\circ}15^{'}  & 2^{\circ}15^{'}  & 0.0123  & 0.0127(05)  \\
    \hline
    -\pi/10  & -4^{\circ}30^{'}  & 4^{\circ}34^{'}  & 0.0489  & 0.0511(11)  \\
    \hline
    -3\pi/20 & -6^{\circ}45^{'}  & 6^{\circ}57^{'}  & 0.0109  & 0.1096(15)  \\
    \hline
    -\pi/5   & -9^{\circ}        & 9^{\circ}29^{'}  & 0.191   & 0.1889(19)  \\
    \hline
    -\pi/4   & -11^{\circ}15^{'} & 12^{\circ}14^{'} & 0.293   & 0.2883(23) \\
    \hline
    -3\pi/10 & -13^{\circ}30^{'} & 15^{\circ}19^{'} & 0.412   & 0.4092(24) \\
    \hline
    -7\pi/20 & -15^{\circ}45^{'} & 18^{\circ}54^{'} & 0.546   & 0.5340(24) \\
    \hline
    -2\pi/5  & -18^{\circ}       & 23^{\circ}18^{'} & 0.691   & 0.6812(22)  \\
    \hline
    -9\pi/20 & -20^{\circ}15^{'} & 29^{\circ}21^{'} & 0.844   & 0.8367(17) \\
    \hline
    -\pi/2   & -22^{\circ}30^{'} & 45^{\circ}       & 1.000   & 0.9951(03) \\
    \hline
\end{array}
$$
\caption{\footnotesize{$\theta $, the angle related to the input states $\psi_{-}$ which is prepared by HWP1 rotated to $\theta_{1/2}$; $C_{1}^2$, the angle of HWP2 to realise the operator $C_{1}^2$. $P_{theory}$ and $P_{e}$ represent the theoretical and the experimental successful probability.}}
\end{table}

\end{document}